# A Simple Network Management Architecture for Supporting Network Administrator and QoS Requirements

Anan Phonphoem and Aphirak Jansang
Department of Computer Engineering, Faculty of Engineering,
Kasetsart University, Bangkok 10900, Thailand
Email: anan@cpe.ku.ac.th; g4365035@ku.ac.th

**Abstract**

*In this paper, a simple network management architecture for supporting both QoS requirements and organization network management policies is purposed. By grouping the traffic flows according to the QoS requirements or certain network management policies, the network resources are effectively controlled. The purposed architecture is easy to deploy; the gateway is the only equipment that needs installation, leaving the rest of the system untouched. The architecture has not significantly degraded the overall system utilization when applying it to the outgoing bound of the gateway. The architecture can also be implemented on the wireless LAN at the access point because the architecture is designed in such the way that it is independent to both the lower and upper protocol layers.*

## 1.0 Introduction

Recently, the Internet saw a dramatic increase on the demand for bandwidth to service a new class of traffic: Multimedia. The most obvious examples of such traffic are videoconferencing, collaborative working, remote teaching and learning, and live broadcasts. Unlike the best effort traffic, delay tolerant traditional Internet, multimedia consumes enormous amount of bandwidth, requires real-time response, and is sensitive to both delay and delay variation. The details of the multimedia can be referred to Multimedia Communication Forum (MMCF) [1]. Unfortunately, multimedia traffic is not the only type of applications that utilize the network resources. Regular Internet traffics such as file transfer, email and web serving, are also competing for the same finite network resources.

In this context, much effort and considerations have been placed upon managing different traffic types and requirements. Quality of Service (QoS) has been introduced to serve as a guideline for managing the network resources. It is used to describe the level of services required by certain applications such as the bandwidth requirement, acceptable delay and security level. The network management person will implement some policies that grant more network resources to the high QoS requirement applications while not starving the rest.

To successfully meet the required QoS, applications and services need to be grouped based on their QoS requirements. As the huge growing demand of the current Internet, the Internet Engineering Task Force (IETF) [2] has formed an Integrated Service Working group to categorize services into several classes. The details can be found at [3].

Once the grouping has been done, a number of alternate queues with certain QoS requirements are assigned for each group. To successfully guarantee the QoS for an application, End-to-end QoS guarantees must be achieved. The QoS commitments are needed to apply on each network node or router, which data packets traverse from source to the destination.

Besides that, an organization might want to control the network resources based on certain organization requirements. For example, the organization key person should have more priority to utilize network resources than workers in the production line that are running same applications. Some departments or groups of IP addresses should have more priority than the others. Thus, the management system must also be flexible enough to accommodate broad spectrum of policies.

In this paper, a simple network management architecture for supporting both QoS requirements and organization network management policies is purposed. The architecture can be implemented on either the ingress or egress of the network. Normally an organization owns the gateway and has the authority to modify the device. Hence, the LAN gateway or a wireless access point becomes the suitable location for implementing QoS and network resource management policies. In conjunction with some management protocols implemented on the Internet, such as the Resource Reservation Protocol (RSVP) [4,9], the end-to-end QoS guarantee can be achieved.

For simplicity without scarifying the ability to manage, traffic is classified into two main categories according to their QoS requirements: Real time and Non-Real time. Multimedia falls into the real time category while the other is categorized as the non-real time application. Hence, two separate queues are created. One is the real time queue; the other is the non-real time queue.

In the next section, network resource management grouping policy has been discussed. In section 3, the simple network management architecture framework is purposed. Section 4 describes the testbed and testing methods. The test results have been shown on section 5. Then, section 6 concludes the paper.

**2.0 Network resource management grouping policy**

Currently, media access control protocols are based on Carrier Sense Multiple Access (CSMA) method. While stations on the same LAN segment are competing for accessing the media, real time and non-real time traffic have equal opportunity to grab the network resources. Even though both types of applications are different in their characteristics, they are treated the same. Without implementing any policy, especially for the real time applications, each data packet will not experience the same delay. Real time applications will then suffer from the unbounded delay and delay variations.

Many researchers [3,4,6,7,8] purposed methods to support QoS by implement some policies to the centralized polling based system [11,12,13,14]. The system uses a central controller, called arbiter or master station, to arbitrate media. An arbiter carefully controls the media by granting the permission to other stations in the system according to their QoS requirements. With this method, the bandwidth and packet delay are guaranteed and QoS requirements can be achieved. However, the overall system bandwidth utilization is decreased due to the polling mechanism overhead especially in the light and medium traffic load.

To be able to control the network resources, traffic must be grouped based on their QoS requirements. However, to control the network resources based on a certain organization policy, the grouping must be done differently. Three grouping methods are suggested: 1) grouped by application type, 2) grouped by IP address or department which the station is located, and 3) grouped by both application and IP address. After grouping has been done, each group will be assigned a percentage based on its priority. A high priority group will receive a high percentage.

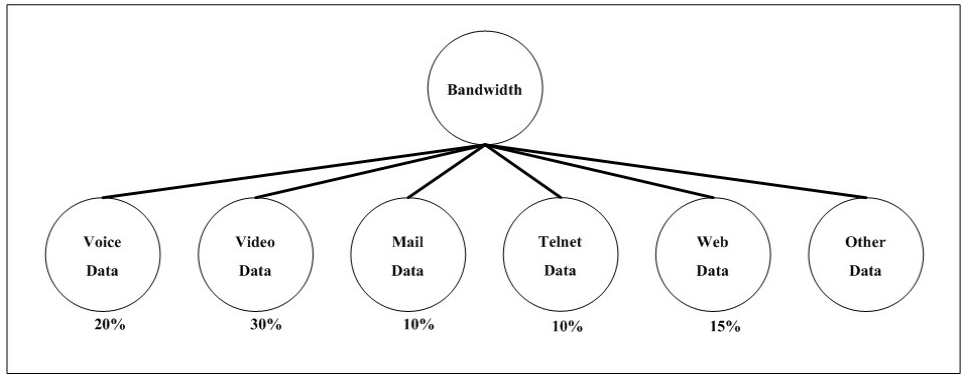

**Figure 1:** Bandwidth allocation grouped by department or IP address

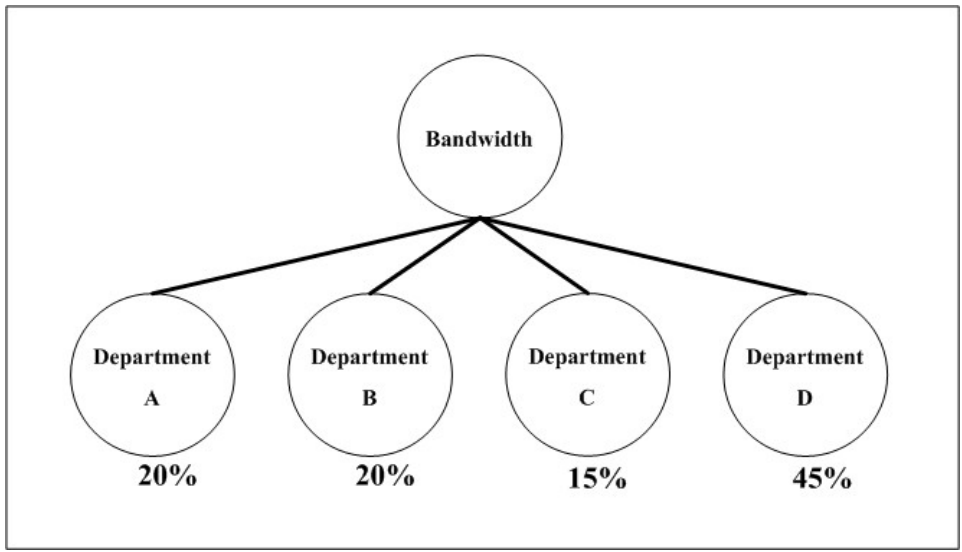

**Figure 2:** Bandwidth allocation grouped by the application type

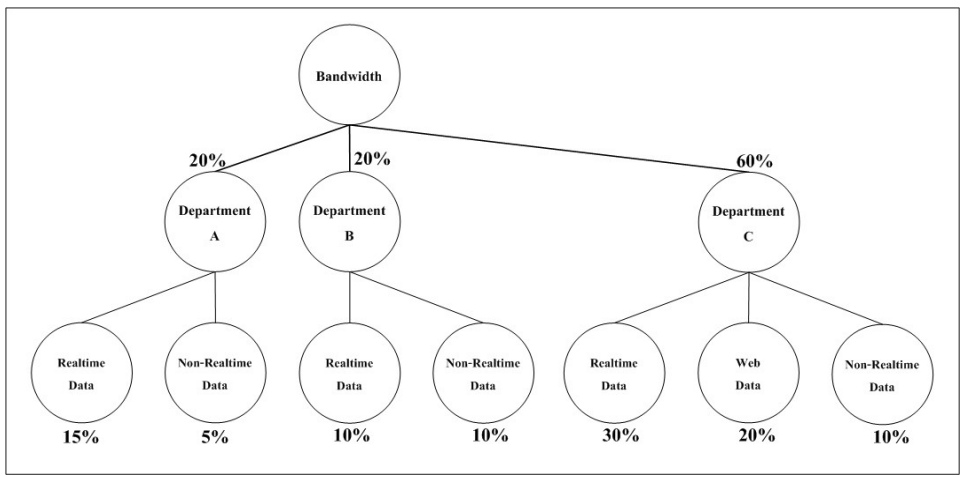

**Figure 3:** Bandwidth allocation grouped by either department or IP address and application type

Figure 1 depicts the example of bandwidth allocation assignment based on its application type. We can prioritize the system by applying a higher percentage to more important application (real time traffic), while assign lower percentage to less important applications (non-real time traffic).

Figure 2 shows the example of bandwidth allocation assignment based on either the department or IP address. We can assign the highest percentage to the most important department and assign less percentage to others. This grouping method might not be able to guarantee the QoS support for each traffic flow. However, it fulfills some management requirements.

Figure 3 displays the example of bandwidth allocation assignment based on both application type and either IP address or department. This grouping method allows us to support the required QoS while be able to assign more priority to certain group of people or departments.

**3.0 The simple network management architecture**

We purpose a simple network management architecture which can be easily implemented on the network gateway. Hence, the other stations in the network segment needs not to be modified or aware of this implementation. With our design, the upper (IP and above) and lower (MAC and below) protocol layers of the gateway machine are untouched. The gateway machine can be able to support any application and works with any network card.

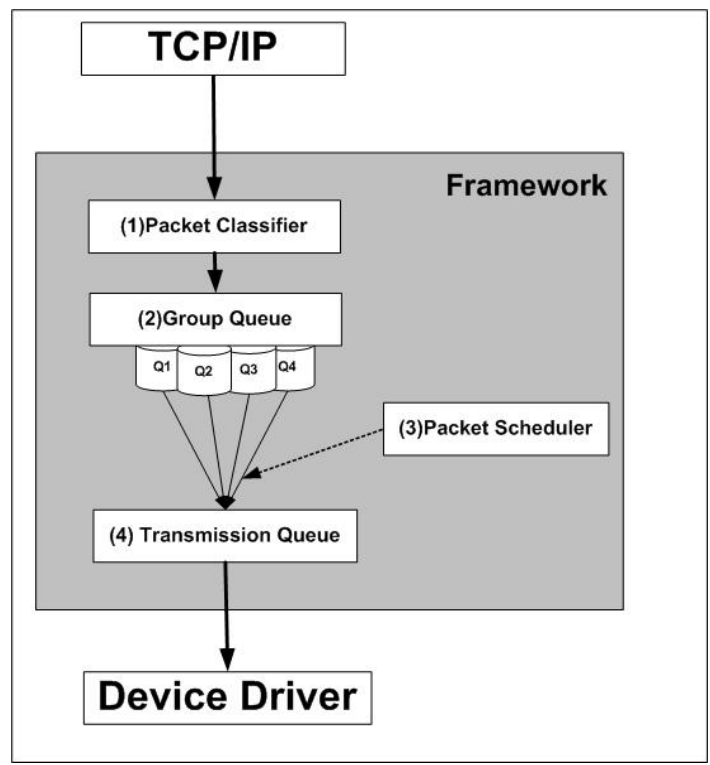

**Figure 4:** A Simple Network Management Architecture

As shown in figure 4, the purposed architecture is composed of four major components:

- **Packet Classifier** – the packet classifier investigates the header of each data packet. It then decides which group that the packet belongs. Finally appends it to the appropriate queue. In case the grouping by application is being applied, the packet classifier categorizes the packet based on well-known port numbers. In case of grouping by department or IP address is used, the packet classifier categorizes the packet based on its source address.
- **Group Queue** – each queue represents the assigned group. All queues are independent and implemented FIFO. Each queue must have a suitable queue size to diminish the packet lost.
- **Packet Scheduler** – the packet scheduler responses for selecting packets from the Group Queue and append them to the Transmission Queue for transmission. The selection process is based on the assigned percentage for each group. The percentage represents the bandwidth ratio that each group receives.
- **Transmission Queue** – The transmission queue is also FIFO. Each packet entering to this queue will be sent out to the network by the regular transmission process of the system

## 4.0 Testbed and testing method

The goal for setting up the testbed is to verify the proposed simple network management architecture. Regularly, the proposed architecture can be implemented either on the outgoing side, from the LAN to the Internet, or on the incoming side, from the Internet to the LAN. Hence, the best location where the architecture should be implemented is also needed to be investigated.

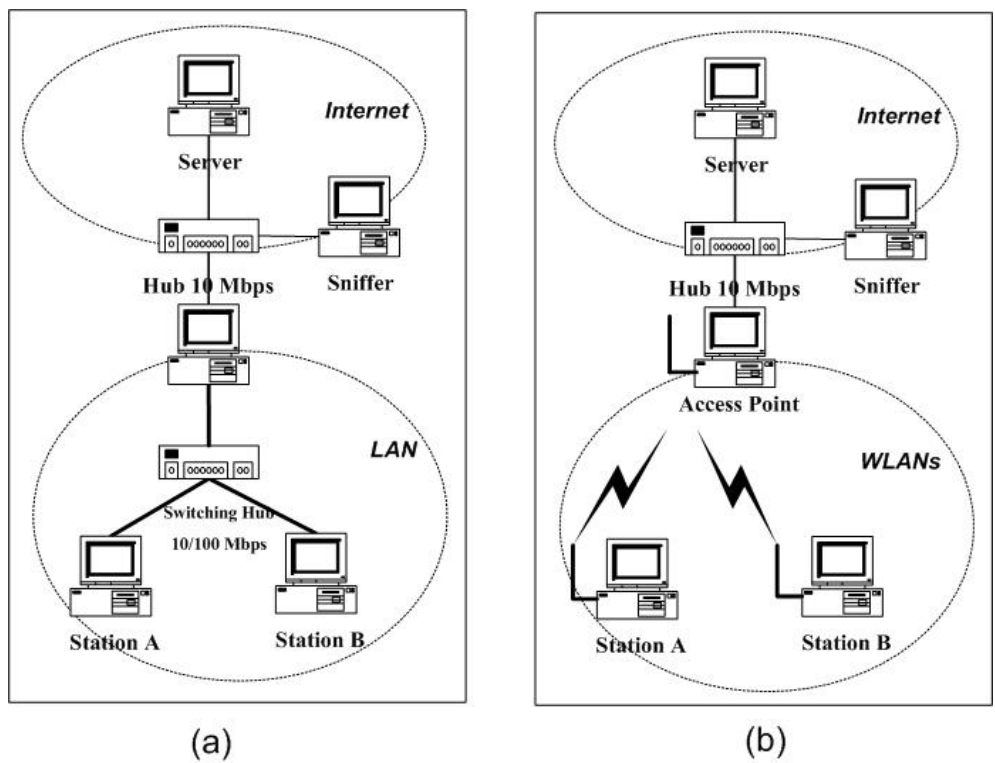

**Figure 5:** The Testbed

Figure 5 depicts our testbed which composes of a gateway or access point, two PC stations represent end users, a server, a sniffer, a 10 Mbps hub, and a 10/100 Mbps switch.

In our experiment, all PC stations are Athlon 700 MHz running Linux operating system kernel version 2.4.16 [5,10]. We select to develop the system architecture on the Linux machine due to its open source characteristics. In fact, only the gateway station that needs to be Linux machine, while the others can run any operating system such as Windows that supports TCP/IP.

The testbed has been setup to represent a regular working condition in an organization. An organization owns a LAN which has a gateway to connect to the Internet via dialup, ISDN, cable modem or leased line. The end users, such as station A and B shown in Figure 5, communicate to servers outside the LAN. At a given time, users might run real time or non-real time applications without knowing the status of the others.

Four experiments are setup as follows:

**Experiment 1:** *Test for the controllability of the architecture by grouping the traffic based on its IP address*

    **Part I**: Implement the architecture on the outgoing bound (Figure 6a)
        1.1 Setup the testbed as shown in Figure 5a
        1.2 Sniff the packet for analysis while running ftp from station A and B to the server
        1.3 Activate the simple network management architecture
        1.4 Assign the percentage X = 50 % to station A and (100-X)% to station B
        1.5 Sniff the packet for analysis while running ftp from station A and B to the server
        1.6 Vary the percentage X to 60, 70, 80 and 90 % and retest using test procedure 1.4

    **Part II**: Repeat steps in Part I but implement the architecture on the incoming bound (Figure 6b)

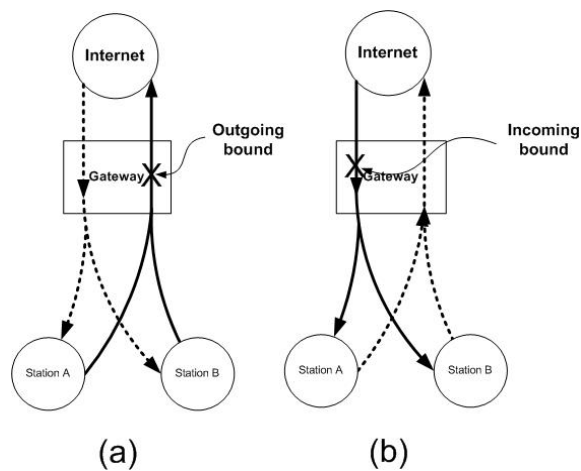

(a)        (b)

**Figure 6:** Experiment 1 configuration

**Experiment 2:** *Test for the controllability of the architecture by grouping the traffic based on applications*

    **Part I**: Implement the architecture on the outgoing bound (Figure 7a)
        2.1 Setup the testbed as shown in Figure 5a

2.2 Sniff the packet for analysis while running a web application from station A (represents real time application) and a running ftp from station B to the server (represents non-real time application)
2.3 Activate the simple network management architecture
2.4 Assign the percentage (X = 50 %) to web application, and (100-X)% to ftp application
2.5 Sniff the packet for analysis while running web application from station A (represents real time application) and running ftp from station B to the server (represents non-real time application)
2.6 Vary the percentage X to 60, 70, 80 and 90 % and retest using test procedure 2.4.

Part II: Repeat steps in Part I but implement the architecture on the incoming bound (Figure 7b)

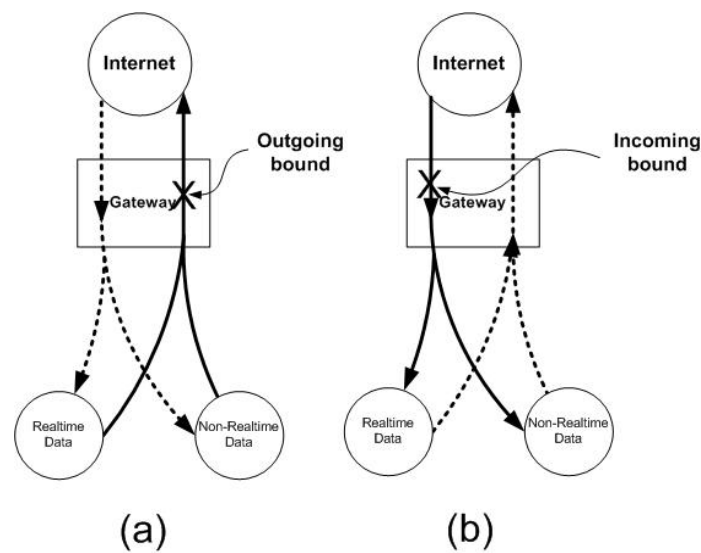

**Figure 7:** Experiment 2 configuration

**Experiment 3:** *Test for the QoS support by grouping the traffic based on applications by implementing on the outgoing bound (Figure 8)*

3.1 Setup the testbed as shown in Figure 5a
3.2 Activate the simple network management architecture
3.3 Assign the fixed percentage X = 50 % to a web application, and (100-X) to a ftp application
3.4 Sniff the packet for analysis while running a web application from station A (represents real time application) and running ftp from station A and B to the server (represents non-real time application)
3.5 Increase the ftp session on station A and B one at time and retest using test procedure 3.4

**Experiment 4:** *Test for wireless LAN implementation*

Repeat the same experiment 1 part I (Figure 6a) by changing from wired LAN (Ethernet) to wireless LAN (IEEE 802.11b) shown in Figure 5b.

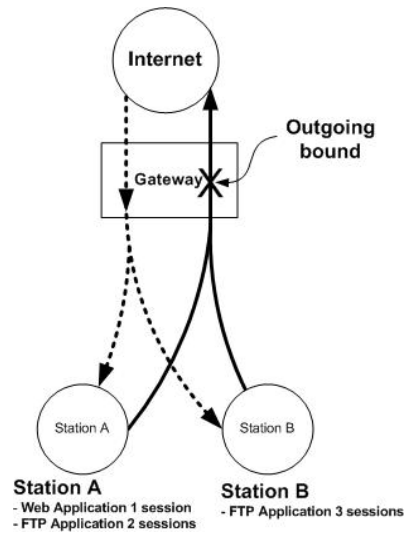

**Figure 8:** Experiment 3 configuration

## 5.0 Test results

From Experiment 1, Part I and II, the network utilization of the system when applying and not applying the architecture are compared as shown in Figure 9. From Experiment 2, Part I and II, the result is as same as the result from Experiment 1. From Experiment 3, the bandwidth utilization of the system when applying fixed bandwidth allocation to the real time application is shown in Figure 10. The bandwidth of the real time application still maintains while all Ftp applications share the rest of the bandwidth. Figure 11 displays the network utilization of the system when applying and not applying the architecture to a wireless LAN in the Experiment 4.

## 6.0 Discussion and Conclusion

In this paper, a simple network management architecture for implementing on the LAN gateway is purposed. By grouping the traffic flows according to the QoS requirements or certain network management policies, the network resources are effectively controlled. The purposed architecture is easy to deploy; the gateway is the only equipment that needs installation, leaving the rest of the system untouched.

From the experiment, the purposed architecture has not significantly degraded the overall system utilization when applying it to the outgoing bound of the gateway. The reason that implementing the architecture on the outgoing bound outperforms implementing it on the incoming bound is the number of packets. The outgoing packets are normally request packets which has a small amount compare with incoming packets, the data from the server. Therefore, the computation and operation time is less for the outgoing bound. This represents the real scenario of the network that users normally download the data more than upload them.

We can also implement the purposed architecture to the wireless LAN on an access point because the architecture is designed in such the way that it is independent to the lower and upper protocol layers.

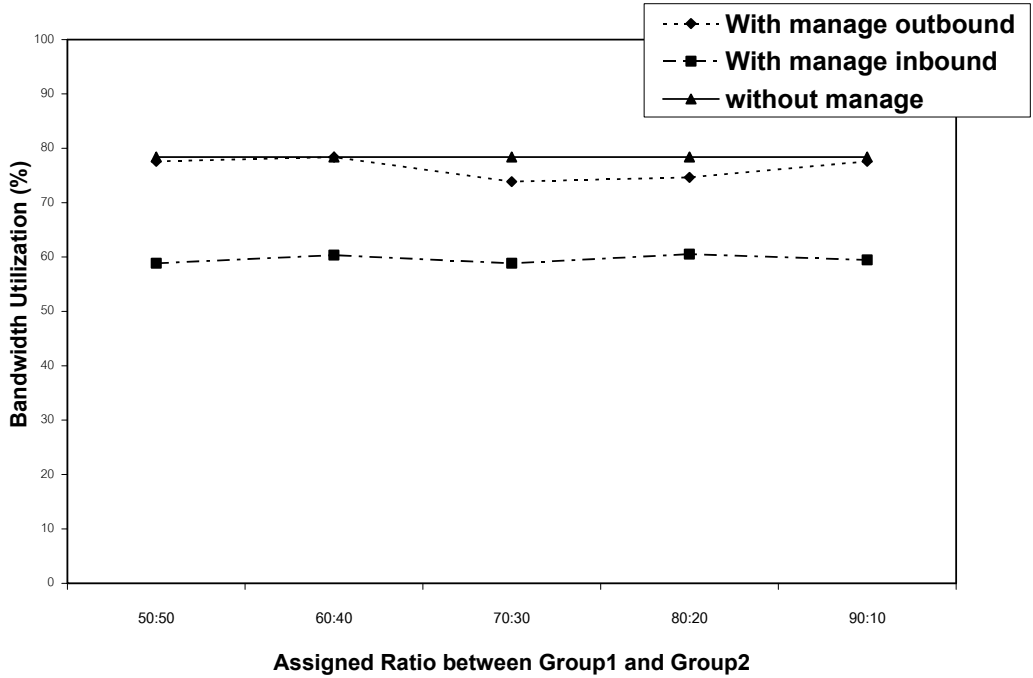

**Figure 9:** The network utilization of the system when applying and not applying the simple network management architecture from wired LAN

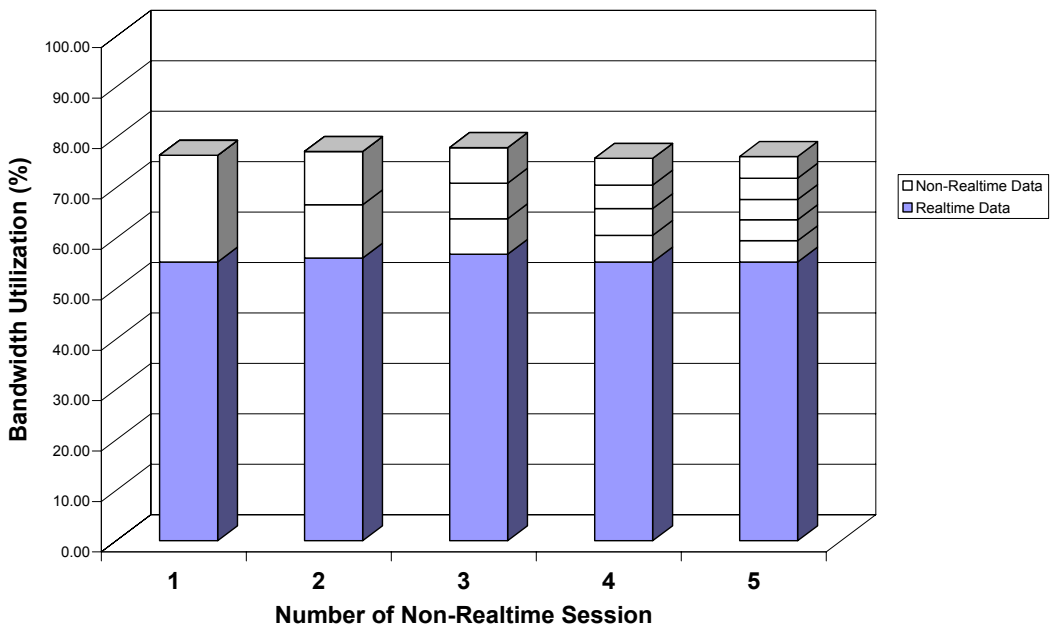

**Figure 10:** The bandwidth utilization of the system when applying fixed bandwidth allocation to real time application and vary the number of non-real time applications

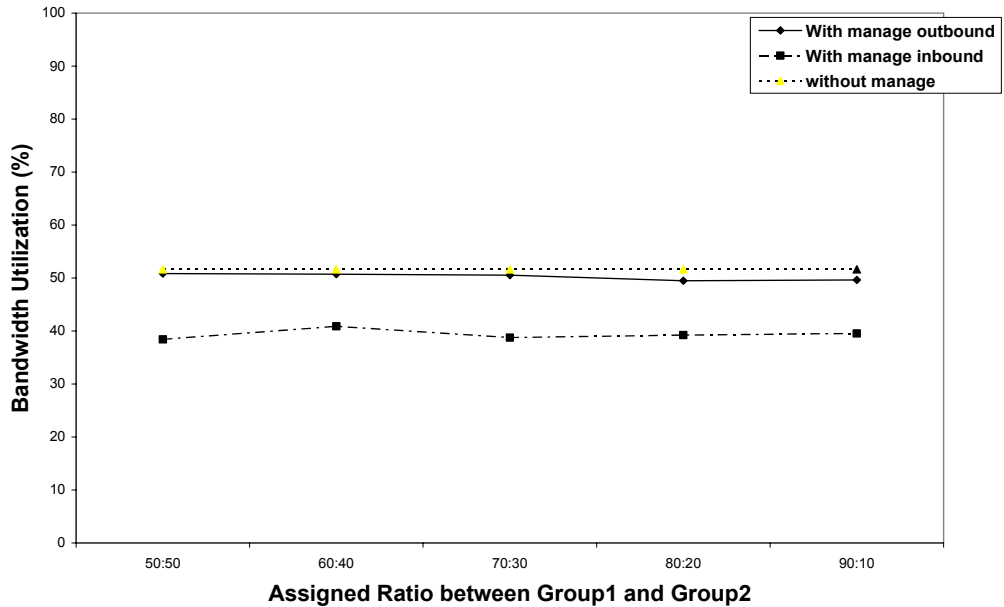

**Figure 11:** The network utilization of the system when applying and not applying the simple network management architecture from the wireless LAN